\documentclass[aps,twocolumn,tightenlines,floatfix,showpacs]{revtex4}

\usepackage{epsfig} 
\usepackage{amssymb} 
\usepackage{amsmath} 
\usepackage{amsfonts}

\newcommand{\be}{\begin{equation}}
\newcommand{\ee}{\end{equation}}

\begin{document}

\title{Tau neutrino and antineutrino cross sections}

\author{Yu Seon Jeong and Mary Hall Reno}
\affiliation{Department of Physics and Astronomy, University of Iowa, Iowa City, IA 52242 USA}

\begin{abstract}
Tau neutrino and antineutrino interactions with nucleons in large underground or under ice detectors will be important signals of astrophysical and atmospheric sources of neutrinos. We present here 
a theoretical update of the deep inelastic scattering 
contribution to the tau neutrino and antineutrino charged current cross sections with isoscalar nucleon targets and proton targets for incident neutrinos and antineutrinos
in the energy range from 10 GeV to 10 TeV. Next-to-leading order quantum chromodynamic corrections, target mass corrections and heavy quark effects are included.
Uncertainties in the cross section associated with the structure functions a low momentum transfers, the  input parton distribution functions, scale dependence and flavor number scheme are discussed.
\end{abstract}


\maketitle

\section{Introduction}

Neutrinos are signals of hadronic interactions that produce mesons, followed by meson decay. There are many models which predict neutrino production in astrophysical sources \cite{ghsreview} as well
as in the atmosphere \cite{gaisser}. IceCube \cite{icecube} and the detector element DeepCore 
\cite{deepcore}
and large underwater detectors \cite{antares}
aim to measure neutrino signals from atmospheric production, from individual sources and the isotropic neutrino flux summed over all the sources. 

The phenomenon of neutrino oscillations is an important ingredient in evaluating the potential signals in detectors. From combined solar and atmospheric
neutrino measurements, we know that the three flavors oscillate into each other \cite{pdg}.
Direct production of tau neutrinos in the atmosphere is small \cite{atmtau}, but oscillations of atmospheric muon neutrinos to tau neutrinos can result in large upward 
$\nu_\tau + \bar{\nu}_\tau$ fluxes at certain energies\cite{leelin,gmm}.
Astrophysical distances are much larger than terrestrial distances, so even if the dominant production in astrophysical sources is
$\nu_\mu+\bar{\nu}_\mu$, a substantial fraction of the
flux will oscillate to $\nu_\tau + \bar{\nu}_\tau$. With the low atmospheric
backgrounds at high energies, tau neutrino signals from convential
and exotic \cite{covi} astrophysical sources may be important. 
The DeepCore
detector may be capable of distinguishing tau neutrino events \cite{gmm}. 

An essential input to theoretical predictions and experimental analyses
of tau neutrino induced events is the tau neutrino charged current
cross section \cite{albright}.
We provide here an update of earlier work \cite{krtau,proc} with a summary of tau neutrino and antineutrino deep inelastic scattering (DIS) charged-current cross sections in the energy range of 10 GeV - 10 TeV. Tau mass corrections, relative to the muon neutrino
charged current cross section, are important for some of this energy range.
The low $Q^2$ extrapolations of the nucleon structure functions,
not included in Refs. \cite{krtau,proc}, are also relevant to the low end of the energy range. We evaluate the next-to-leading order quantum chromodynamic
corrections \cite{charm}, the effect of target mass 
corrections \cite{schienbein,kr,georgi}
and the range of predictions from different ways of
incorporating the charm quark contribution \cite{aot,acot,leshouches}. 
A new estimate of the theoretical errors for the DIS cross sections for tau neutrino and antineutrino charged-current interactions is presented.

\section{Cross section for tau neutrinos}

In the lower end of the energy range discussed here, target mass 
\cite{georgi,schienbein,kr},
tau mass \cite{albright,krtau} and quark mass \cite{charm}
corrections are important \cite{proc}.
The low momentum transfer behavior of the structure functions is also a consideration. To discuss these corrections, we define the
integration variables $x\equiv Q^2/2 p\cdot q$ and $y\equiv p\cdot q/p\cdot k$,
where the momentum assignments are
\begin{eqnarray}
\nonumber \nu/\bar{\nu}
_\tau (k)\ &+&\ N(p) \rightarrow \tau/\bar{\tau}(k') \ +\  X \\
q^2&\equiv &(k-k')^2 = -Q^2\ .
\end{eqnarray}

The target nucleon mass $M$ comes in primarily
through the replacement of the parton light cone momentum fraction
$x$ by the Nachtman variable $\eta=x(2/(1+\sqrt{1+4M^2x^2/Q^2}\,))$.
Target mass corrections (TMC) also modify the structure functions themselves, where
the uncorrected functions are denoted by $F_i$ and the target 
mass corrected structure functions are $F_i^{\rm TMC}$. The difference between
$F_i$ and $F_i^{\rm TMC}$ is largest at large $x$.
For a summary of target mass corrections to the structure functions, see, for example, Refs. \cite{schienbein,kr,georgi}.

The tau mass $m_\tau$ is kept explicitly in the differential
cross section. By keeping terms proportional to
$m_\tau^2/ME_\nu$, five of the six target mass
corrected structure functions allowed by
the general Lorentz structure of the hadronic current appear in the 
differential cross section \cite{albright}. 
The expression for the tau neutrino charged current differential cross section, for deep inelastic scattering (DIS), is
\begin{eqnarray} \nonumber
\frac{d^2\sigma^{\nu(\bar{\nu})}}{dx\ dy} &=& \frac{G_F^2 M
E_{\nu}}{\pi(1+Q^2/M_W^2)^2}
\Biggl(
(y^2 x + \frac{m_{\tau}^2 y}{2 E_{\nu} M})
F_1^{\rm TMC} \\ \nonumber
&+& \left[ (1-\frac{m_{\tau}^2}{4 E_{\nu}^2})
-(1+\frac{M x}{2 E_{\nu}}) y\right]
F_2^{\rm TMC}
\\ \nonumber
&\pm& 
\left[x y (1-\frac{y}{2})-\frac{m_{\tau}^2 y}{4 E_{\nu} M}\right]
F_3^{\rm TMC} \\  
&+& 
\frac{m_{\tau}^2(m_{\tau}^2+Q^2)}{4 E_{\nu}^2 M^2 x} F_4^{\rm TMC}
%
- \frac{m_{\tau}^2}{E_{\nu} M} F_5^{\rm TMC}
\Biggr)\, ,
\label{eq:nusig}
\end{eqnarray} 
where neutrino (antineutrino) scattering requires $+F_3^{\rm TMC}$ ($-F_3^{\rm TMC}$) in
eq. (\ref{eq:nusig}).  

For muon neutrino charged current interactions, $F_4^{\rm TMC}$ and $F_5^{\rm TMC}$
are generally omitted since they are suppressed by a factor of at least
$m_\mu^2/2ME_\nu$ relative to the contributions of $F_1$, $F_2$ and
$F_3$.
Albright and Jarlskog noted that at leading order, in the massless quark
and massless target
limit, $F_4=0$ and $F_5=F_2/2x$ \cite{albright}. These are called the Albright-Jarlskog (AJ) relations. 
Corrections to these relations at next-to-leading order (NLO) in quantum chromodynamics including
charm mass corrections appear, for example,
in Ref. \cite{krtau}.

A lepton mass correction to the charged current
cross section comes from the kinematic limits on $x$ and $y$
\cite{albright,krtau}. For completeness,
we reproduce them here:
\begin{eqnarray}
 && \frac{m_\tau^2}{2M(E_\nu -m_\tau)}\ \leq \  x\ \leq\ 1\ ,\\
&&a\ -\ b\ \leq \ y\ \leq \ a\ +\ b \ ,
\end{eqnarray}
where the quantities $a$ and $b$ are
\begin{eqnarray*}
a & = & \Biggl[1-m_\tau^2\Biggl(\frac{1}{2ME_\nu x}+\frac{1}
{2E_\nu ^2}\Biggr)\Biggr]  
/(2+Mx/E_\nu)\ ,\\
b & =&  \Biggl[\Biggl(1-\frac{m_\tau^2}{2ME_\nu
  x}\Biggr)^2-\frac{m_\tau^2}{E_\nu ^2}\Biggr]^{1/2}  
/(2+Mx/E_\nu )\ .
\end{eqnarray*}

A final kinematic issue relates to the DIS nature of the scattering.
For quasielastic scattering, e.g., $\nu_\tau n\rightarrow \tau p$,
the structure functions are proportional to the delta function 
$\delta (W^2-M^2)$ where $W^2$ is the invariant mass of the
hadronic final state. These multiply the nucleon form factors
\cite{llsmith,sv}. Nuclear physics models are used to evaluate few
pion production \cite{fewpion,fewpionnew}. To avoid double counting these
exclusive contributions in our DIS evaluation, we require
that the invariant mass of the hadronic final state $W$ be larger
than a minimum value. Our standard choice is $W_{\min} = 1.4$ GeV.
In terms of $x$ and $Q^2$, this means
\begin{equation}
W^2=Q^2\Bigl(\frac{1}{x}-1\Bigr)+M^2\geq W^2_{\rm min}\ .
\end{equation}

The minimum $W^2$ as well as the tau mass come into play in the differential distribution in $y$. This
is most conveniently evaluated using
\begin{equation}
\frac{d\sigma}{dy} = \int_{Q^2_{\rm min}}^{Q^2_{\rm max}}dQ^2\, \frac{d\sigma}{dy\, dQ^2}\ ,
\end{equation}
where the limits of integration are
\begin{eqnarray*}
Q^2_{\rm min} & = & 2 E^2(1-\epsilon)(1-y)-m_\tau^2\ ,\\
Q^2_{\rm max} &=& 2MEy + M^2-W^2_{\rm min} = 2MEy - \Delta^2  \ ,
\end{eqnarray*}
where $\epsilon = \sqrt{1-m_\tau^2/\Bigl( (1-y)E\Bigr)^2}$ and $\Delta^2 = W^2_{\rm min}-M^2$.
Now $y$ ranges between
\begin{eqnarray*}
&& y_{\rm min} = \frac{1}{4ME}\times\Biggl( 2ME + \Delta^2-m_\tau^2\\
& -& \sqrt{(2ME-\Delta^2)^2+m_\tau^4-2(2ME+\Delta^2)m_\tau^2}\Biggr)\ ,\\
&&y_{\rm max} = 1-m_\tau/E\ .
\end{eqnarray*}

Even with $W^2>W^2_{\rm min}$, $Q^2$ can be below the nominal minimum
$Q^2$ for the parton distribution functions (PDFs) used to evaluate
$F_i^{\rm TMC}$. We use a 
phenomenological parameterization of the
structure functions first discussed by Capella, Kaidalov, Merino and Tran Than (CKMT)
\cite{ckmt} and later used in Ref. \cite{reno} for the low $Q^2$ extrapolations needed 
in $\nu_\mu N$ scattering. The CKMT parameterization of the
structure functions at low $Q^2$ accounts for the nonperturbative
behavior of $F_i^{\rm TMC}$. The CKMT parameterization is matched at
$Q^2=Q_c^2$ to the perturbative evaluations
of $F_i^{\rm TMC}$ with PDFs. Below this cutoff momentum transfer $Q_c$, 
we use the AJ relations for $F_4^{\rm TMC}$
and $F_5^{\rm TMC}$, regardless of the incident
neutrino energy. In our results below, we exhibit the sensitivity of the cross
section to different choices of $Q_c$. These results are quite similar to those using the
Bodek-Yang-Park prescription \cite{bodek} for the low $Q^2$ extrapolation of the structure
functions \cite{reno}.

While we use the AJ relations for low $Q^2$, we can test the AJ relations for
higher $Q^2>Q_c^2$ values. Setting $F_4^{\rm TMC}=0$ for all $Q^2$ results in
a neutrino cross section only 1\% higher than for $F_4^{\rm TMC}\neq 0$ at 10
GeV, while for antineutrinos, the enhancement is 4\%. At 100 GeV, the 
$F_4^{\rm TMC}$ contribution is negligible.  By substituting $F_5^{\rm TMC}
=F_2^{\rm TMC}/2x$ at 100 GeV, the cross sections change by less than 1\%.
They differ by 2\% for $\nu_\tau N$ and by 9\% for $\bar{\nu}_\tau N$ at
10 GeV as compared to the full calculation of $F_5^{\rm TMC}$ in Ref. \cite{kr}. We use
the full NLO calculation with TMC of Ref. \cite{krtau,kr} for $E_\nu\leq 100$ GeV
and $Q^2\geq Q_c^2$. The AJ relations
for $E_\nu>100$ GeV reproduce the full calculations of the cross sections.

\begin{table}
\label{table:sigma}
\begin{tabular}{|l|c|c|}
\hline 
Energy [GeV] & $\sigma_{\nu N}\ [10^{-38}\ {\rm cm}^2]$ & $\sigma_{\bar{\nu} N}\ [10^{-38}\ {\rm cm}^2]$ \\
\hline\hline
10 &  $0.916\ (1.26,\ 0.690)$	& $0.291\ (0.574,\ 0.160)$	\\
\hline
$10^{1.25}$& 3.77	(4.22,\ 3.44)&	1.48 (1.90,\ 1.21)\\
\hline
$10^{1.5}$&	10.4 (10.9,\ 9.97)&	 4.43 (4.93,\ 4.05)\\
\hline
$10^{1.75}$&23.7 (24.3,\ 23.2)	&	10.6 (11.2,\ 10.2)\\
\hline
$10^2$& 48.9 (49.5,\ 48.4)	&	22.8 (23.4,\ 22.3)\\
\hline
$10^3$&	$5.69\times 10^2$&	$3.02\times 10^2$\\
\hline
1$0^4$& $4.30\times 10^3$	&	$2.76\times 10^3$\\
\hline
\end{tabular}
\caption{The $\nu_\tau$-isoscalar nucleon  and $\bar{\nu}_\tau$-isoscalar nucleon
charged current cross section using the CKMT parameterization for 
$Q^2\leq 2$ GeV$^2$ in the structure functions, matched to the perturbative structure functions with GJRF PDFs \cite{gjr}, using the charm mass corrections from
Ref. \cite{kr} and target mass corrections from Ref. \cite{kr,schienbein}. Here, the factorization scale is set to $\mu=Q$, and the hadronic invariant mass is limited to $W\geq 1.4\ {\rm GeV}\ (M,1.7\ {\rm GeV})$ in the
cross section.}
\end{table}

\begin{table}
\label{table:sigmap}
\begin{tabular}{|l|c|c|}
\hline 
Energy [GeV] & $\sigma_{\nu p}\ [10^{-38}\ {\rm cm}^2]$ & $\sigma_{\bar{\nu} p}\ [10^{-38}\ {\rm cm}^2]$ \\
\hline\hline
10 &  $0.472\ (0.646,\ 0.356)$	& $0.428\ (0.840,\ 0.230)$	\\
\hline
$10^{1.25}$& 2.12	(2.36,\ 1.94)&	2.08 (2.69,\ 1.69)\\
\hline
$10^{1.5}$&	6.21 (6.49,\ 5.98)&	 6.04 (6.78,\ 5.51)\\
\hline
$10^{1.75}$&14.8 (15.1,\ 14.6)	&	14.2 (15.0,\ 13.5)\\
\hline
$10^2$& 31.6 (32.0,\ 31.4)	&	29.8 (30.6,\ 29.1)\\
\hline
$10^3$&	$3.99\times 10^2$&	$3.72\times 10^2$\\
\hline
1$0^4$& $3.28\times 10^3$	&	$3.26\times 10^3$\\
\hline
\end{tabular}
\caption{The $\nu_\tau$-proton  and $\bar{\nu}_\tau$-proton, as in Table I, for the DIS
charged current cross section.  The hadronic invariant mass is limited to $W\geq 1.4\ {\rm GeV}\ (M,1.7\ {\rm GeV})$ in the
cross section.}
\end{table}

\section{Results}

For the cross sections evaluated here we have used two different sets of
parton distribution functions. The Gluck, Jimenez-Delgado and Reya  (GJR) PDFs
\cite{gjr} for a fixed flavor number scheme with three
light flavors is used. One feature of the GJR PDFs is that the fits
are done for $Q^2\geq 0.5$ GeV$^2$. This offers the option for a range of
cut-off scales $Q_c^2$, below which the phenomenological CKMT parameterizations 
are used \cite{ckmt,reno}. Our standard choice is $Q_c^2=2$ GeV$^2$. We also
use the CTEQ6.6M PDFs \cite{cteq6p6}. This set includes 44 variations of the best
fit parameter values to help characterize the error associated with the PDF fit.

\vskip 0.5in
\begin{figure}[t]
\includegraphics[width=2.5in,angle=270]{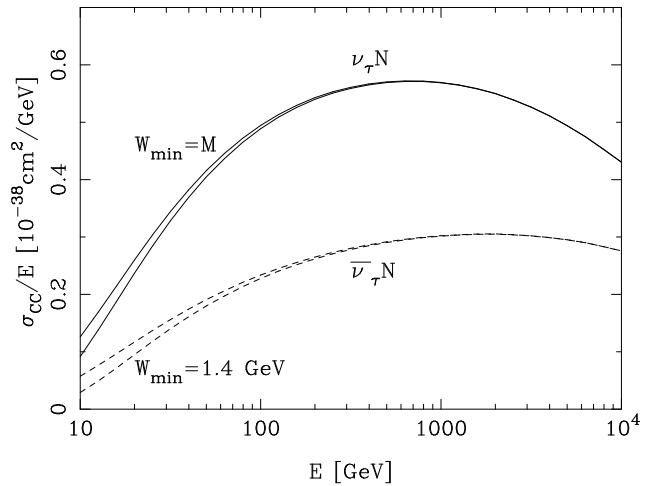}
\caption{The tau neutrino- (solid) and antineutrino- (dashed)
isoscalar nucleon charged current cross section, for
$W>1.4$ GeV (lower curves) and $W>M$ (upper curves). 
Target mass corrections, NLO QCD and low $Q$ extrapolations
of the structure functions below $Q^2_c=2$ GeV$^2$ are included in the
evaluation, as in Table 1.}%
\label{fig:sigma}%
\end{figure}

The charged current cross sections divided by incident tau
neutrino or antineutrino energy with $Q_c^2=2$ GeV$^2$ and the
three flavor GJR PDFs with the factorization scale in the structure functions
set to $\mu^2= Q^2$, are plotted in Fig. 1 for iso-scalar nucleon targets. The solid lines show $\sigma_{CC}(\nu_\tau N)/E$ and
the dashed lines show $\sigma_{CC}(\bar{\nu}_\tau N)/E$. The cross sections for a few
energies are also listed in Table 1. The upper curves show the cross sections when
$W_{\rm min}=M$, while the lower curves have $W_{\rm min}=1.4$ GeV.
Unless otherwise noted, our default evaluation is with $W_{\rm min}
=1.4$ GeV.
Table 1 also lists results when $W_{\rm min}=1.7$ GeV. For completeness, we include in Table 2 the tau neutrino-proton
and tau antineutrino-proton DIS charged-current cross sections.

The rise in the cross section divided by incident neutrino energy at low incident energy is largely from the kinematic
threshold effect. The additional structure function $F_5^{\rm TMC}$, which comes in with a minus sign, has a numerical
effect even at 100 GeV incident energy. Neglecting $F_5^{\rm TMC}$ leads
to an error of 13\% for tau neutrinos and 29\% for tau antineutrinos for $E_\nu=100$ GeV,
with even larger errors at lower energies. At higher energy, W boson propagator already has an effect. For example, the neutrino cross section decreases by 6\% for $E=10^3$ GeV and by 30\% at $E=10^4$ GeV relative to the evaluation with a simple four-Fermi interaction.

\vskip 0.5in
\begin{figure}[t]
\includegraphics[width=3.0in]{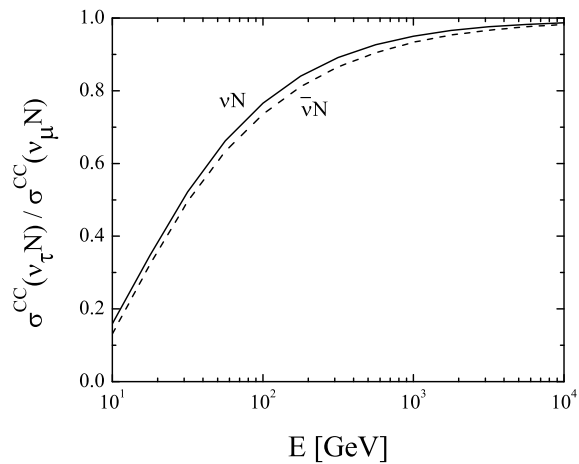}
\caption{The ratio of $\sigma_{CC}(\nu_\tau N)/\sigma_{CC}(\nu_\mu N)$ neutrino- (solid) and antineutrino- (dashed)
DIS cross sections, for
$W>1.4$ GeV.}%
\label{fig:ratio}%
\end{figure}

Overall, the
kinematic suppression of tau production and the structure function term  in the differential cross section
lead to tau neutrino and antineutrino cross sections lower than the muon neutrino cross section over a surprisingly wide
range of energies. With the combined corrections from the structure functions and the kinematic limits, the tau neutrino and antineutrino charged current cross sections
are less than the corresponding muon neutrino cross sections by about 25\%
at 100 GeV incident neutrino energy. Even at 1 TeV, the tau neutrino (antineutrino)
charged current cross sections are reduced by 5\%(7\%) compared to the muon
neutrino and antineutrino cross sections for an isoscalar target.
We show in Fig. \ref{fig:ratio} the ratio of the charged current DIS cross sections for
$\nu_\tau N$ and $\bar{\nu}_\tau N$ scattering for $W_{\rm min}=1.4$ GeV. 

For scattering with a proton target and $W_{\rm min}=1.7$ GeV,
the neutrino ratio is between 4-7\% lower than for what is shown in Fig. \ref{fig:ratio}
for E=10-100 GeV. The antineutrino ratio for a proton target with 
$W_{\rm min}=1.7$ GeV is slightly larger than the ratio in Fig. \ref{fig:ratio}.
At E=100 GeV, it is larger by about 2\% and by $\sim 1\%$ at $10^3$ GeV.

\vskip 0.5in
\begin{figure}[t]
\includegraphics[width=3.0in]{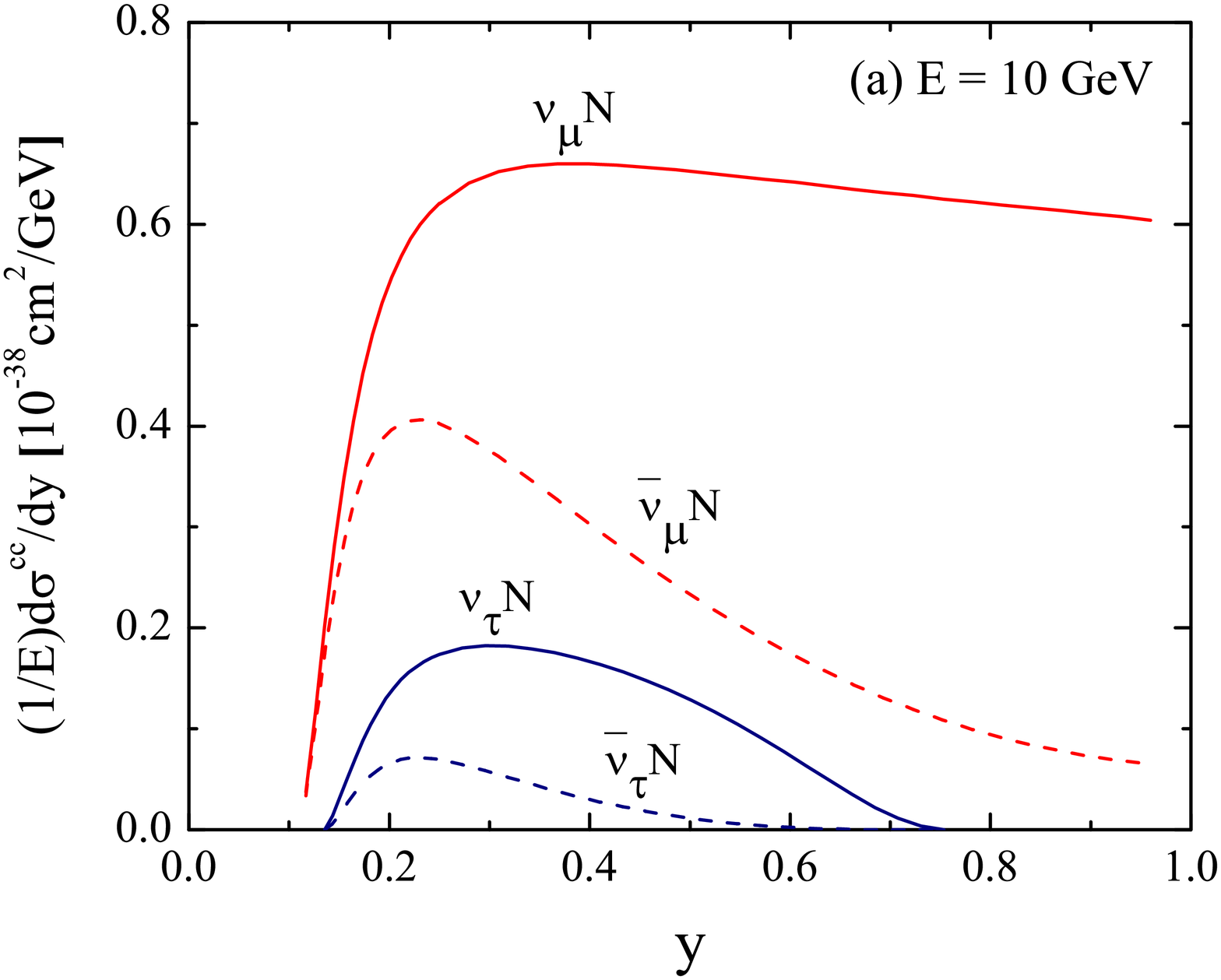}
\includegraphics[width=3.0in]{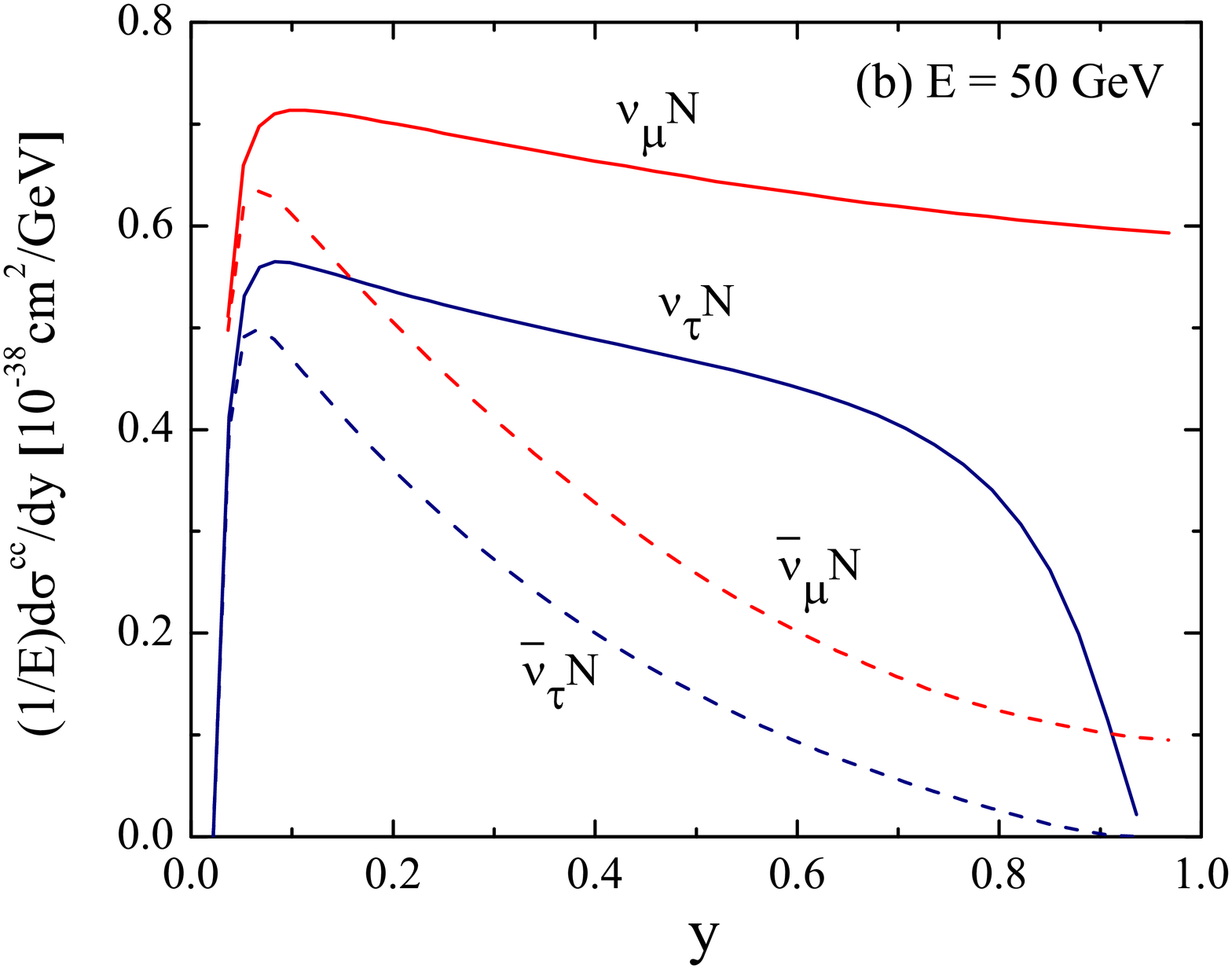}
\caption{The differential cross section $d\sigma/dy/E$ for DIS charged current scattering with an isoscalar nucleon target, with
$W>1.4$ GeV for (a) 10 GeV and (b) 50 GeV incident energies.}%
\label{fig:dsdy}%
\end{figure}

A further demonstration of the kinematic suppression due to tau lepton production is seen in the $y$-distribution. In Fig. \ref{fig:dsdy}, we show $d\sigma/dy/E$ for tau neutrino and antineutrino scattering as well as for muon neutrino and antineutrino scattering for an isoscalar target. We show the differential distribution for two incident energies: for $E=10$ GeV (upper figure) and $E=50$ GeV (lower figure). With muon neutrino scattering, one sees a nearly constant cross section as a function of y, because at these energies, the cross section is valence dominated, so $d\sigma/dy\sim q(x,Q^2)$ for the valence parton distribution function $q(x,Q^2)$. At low $y$, the figure illustrates the consequence of the $W_{\rm min}$ choice. For tau neutrino scattering, the limits in $Q^2$ and $y$ cut off the high $y$ distribution.
For antineutrino scattering, valence domination in the structure functions yields schematically $d\sigma/dy\sim (1-y)^2 q(x,Q^2)$. This
means that the antineutrino differential cross section is already falling with increasing $y$, so the kinematic effects of producing
a tau lepton are less noticible. For both muon antineutrino and tau antineutrino scattering, the low-$Q^2$ range is more emphasized, since
the average $y$ value is lower than for neutrino scattering.

The theoretical error on the DIS cross section has a number of components which we discuss in
the remainder of this section. We begin with the low-$Q^2$ extrapolation using the CKMT structure functions. The GJR PDFs are fit to $Q_0^2=0.5$ GeV$^2$. This allows us to 
compare the cross sections with choices for the transition point between PDFs and the CKMT structure functions $Q_c^2$. We find that for neutrinos, setting $Q_c^2=0.5$ GeV$^2$ results in $\sigma_{CC}(\nu_\tau N)$ to less than 1\% of the cross sections with $Q_c=2$ GeV$^2$ and $W_{\rm min}=1.4$ GeV. As noted, antineutrino scattering occurs at lower values of $Q^2$, so the value of $Q_c^2$ has somewhat more importance for antineutrinos. At $E=10$ GeV, $\sigma(\bar{\nu}_\tau N)$ is
about 3\% lower for $Q_c^2=0.5$ GeV$^2$ than for $Q_c^2=2$ GeV$^2$ with $W_{\rm min}=1.4$ GeV. By $E=100$ GeV, the discrepancy is reduced to $\sim 1$\%.  

The factorization scale dependence of the cross section introduces a larger theoretical error. For
$\nu_\tau N$ scattering, setting the factorization scale to $\mu^2=0.5 Q^2$ enhances the cross section by close to 3\%, while for
$\mu^2 = 2 Q^2$ decreases the cross section by approximately 4\%
as compared to $\mu^2=Q^2$ at $E=10$ GeV. For $\bar{\nu}_\tau N$
scattering, the range at 10 GeV is between $\pm 5\%$
The factorization
scale dependence declines with energy.

\vskip 0.5in
\begin{figure}[t]
\includegraphics[width=2.5in,angle=270]{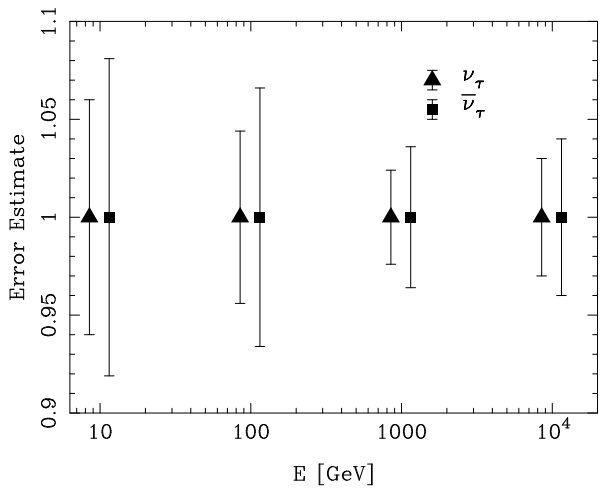}
\caption{An estimate of the theoretical errors for $\nu_\tau$ (triangles) and $\bar{\nu}_\tau$ (squares) deep-inelastic
charged current cross sections for
selected incident (anti-)neutrino energies with $W_{\rm min}=1.4$ GeV.}%
\label{fig:errors}%
\end{figure}

The choice of parton distribution function also has an implication for the predicted DIS charged current cross section. Using 
the CTEQ6.6M PDFs in the evaluation of the cross section using Ref. \cite{kr} yields the $\nu_\tau N$ ($\bar{\nu}_\tau N$)
cross section between 0.6-2.7\% (0.9-3.0\%) higher than using the
three-flavor GJR PDFs for $E=10-10^4$ GeV. 

The CTEQ6.6 PDFs are available with a set of 44 variations of the best fit parameter values. Using these 44 sets, we find that the resulting error in the cross section is largest for the lowest energy. For $\nu_\tau N$ scattering, the PDF uncertainty
is 2.7\% at 10 GeV, reducing to 1.4\% at $10^4$ GeV. Slightly larger for $\bar{\nu}_\tau N$, the estimated CTEQ6.6 PDF error is 3.6\% at 10 GeV and 2\% at $10^4$ GeV.

The final uncertainty in the DIS theoretical prediction considered here
is due to the choice of calculational scheme to include the charm quark.
At $E=10$ GeV, charm contributions are negligible, but as the energy increases, charm contributions become more important.
Beginning at $E=100$ GeV, we have evaluated $\sigma_{CC}(\nu_\tau N)$ and ${\sigma}_{CC}(\bar{\nu}_\tau N)$ using the
ACOT and the S-ACOT-$\chi$ schemes \cite{acot},
which we compare to the fixed flavor scheme result of
Ref. \cite{kr} including massive charm quark corrections.

Using the same CTEQ6.6M PDFs, we see a small range of theoretical predictions for the flavor schemes we used here. 
The S-ACOT-$\chi$ result for $\nu_\tau N$ at
$E=100$ GeV is lower by close to $ 4\%$ than the calculation using Ref.
\cite{kr}, and close to 6\% lower for $\bar{\nu}_\tau N$. 
The discrepancy reduces to close to zero at $E=10^3$ GeV, and
it remains small at $E=10^4$ GeV.

Other
schemes to incorporate charm are possible. A detailed discussion
and comparison of these schemes appears in Ref. \cite{leshouches}. The
ultrahigh energy implications of the scheme dependence in 
neutrino-nucleon scattering are discussed in Ref. \cite{jr1}.

\section{Final Remarks}

We have presented an evaluation of the tau neutrino and tau antineutrino cross sections
for DIS 
charged-current scattering with isoscalar nucleon targets. The Albright-Jarlskog
approximations for $F_4$ and $F_5$ are excellent at high energies, however,
at $E=10$ GeV, for antineutrino scattering, the deviation from the full calculation
of $F_5^{\rm TMC}$ is on the order of $\sim 10\%$ for tau antineutrino
scattering. We have considered the numerical
implications of a range of factorization scales, of the PDF choice and of the cutoff scale $Q_c^2$ below
which a low-$Q$ extrapolation is used for the structure functions. 
We have also made an estimate of the uncertainty of the flavor scheme dependence of the inclusion of the charm quark contribution.

A summary of our estimates of the theoretical uncertainties to the DIS cross
section is shown in Fig. \ref{fig:errors}, where the various contributions have been
added in quadrature.  At 10 GeV, the largest error comes from the scale
dependence of the structure functions, while at the highest energies,
a more important element is the choice of PDF. At intermediate energies,
the flavor scheme (S-ACOT-$\chi$ versus the fixed flavor scheme) for the 
incorporation of charm
gives the largest variation in the theoretical prediction.

Our evaluation of the DIS cross section relies on a minimum value of the final state
hadronic invariant mass $W$. The inclusion of few pion and quasi-elastic neutrino
scattering is required for a full evaluation of the neutrino-nucleon cross section.
At 10 GeV, the quasi-elastic cross section is about 1/3 of the neutrino DIS cross section with $W>1.4$ GeV, and comparable to the antineutrino cross section for the
same $W$ cutoff. The few pion cross sections are lower. Combining the DIS, exclusive
and quasi-elastic contributions in $\nu_\mu$ scattering appears in, e.g., Ref. \cite{naumov}. This procedure is applicable to tau neutrino scattering as well.

The tau neutrino and antineutrino cross sections have been measured
by the DONuT Collaboration, with results summarized in Ref. \cite{donut}.
For an average of tau neutrino and antineutrino cross sections at an
average energy of 115 GeV, 
$$\sigma^{avg}/E = 0.39 \pm 0.13 \pm 0.13 \times 10^{-38} {\rm \ cm}^2 {\rm GeV}^{-1}\ 
$$
for the 9 $\nu_\tau+\bar{\nu}_\tau$ identified interactions. The OPERA
experiment has likely seen its first tau neutrino induced event
from the CERN neutrino beam \cite{opera}.
While the DONuT measurement is the first direct measurement of the tau neutrino 
cross section, one expects that tau neutrino charged current interactions
will play an important role in detailed measurements of neutrino oscillations in
the Earth, for example, with the DeepCore detector in IceCube \cite{deepcore}.

\begin{acknowledgments}
This research was supported by 
US Department of Energy 
contracts DE-FG02-91ER40664. We thank F. Olness for 
providing his computer program to evaluate the quark
mass corrected, NLO structure functions in the various ACOT schemes.
We also thank S. Kretzer for his earlier contributions to the numerical
evaluation of the tau neutrino
cross section.

\end{acknowledgments}



\begin{thebibliography}{99}

\bibitem{ghsreview}
  T.~K.~Gaisser, F.~Halzen and T.~Stanev,
  Phys.\ Rept.\  {\bf 258}, 173 (1995)
  [Erratum-ibid.\  {\bf 271}, 355 (1996)]
  [arXiv:hep-ph/9410384].
  
\bibitem{gaisser}
T.~K.~Gaisser and M.~Honda,
  Ann.\ Rev.\ Nucl.\ Part.\ Sci.\  {\bf 52}, 153 (2002)
  [arXiv:hep-ph/0203272].
  
\bibitem{icecube}
A. Karle, for the IceCube Collaboration, 
in the Proceedings of the 31st ICRC, Lodz 2009,
arXiv:1003.5715.

\bibitem{deepcore}
 C.~Wiebusch, for the IceCube~Collaboration,
  in the Proceedings of teh 31st ICRC, Lodz 2009,
  arXiv:0907.2263 [astro-ph.IM].

\bibitem{antares}
 P.~Kooijman  [ANTARES Collaboration],
  PoS E {\bf PS-HEP2009}, 100 (2009);
 A.~Margiotta  [KM3NeT Collaboration],
  J.\ Phys.\ Conf.\ Ser.\  {\bf 203}, 012124 (2010).
 
\bibitem{pdg}
C. Amsler et al. (Particle Data Group), 
Phys. Lett.  {\bf B667}, 1 (2008). 

\bibitem{atmtau}
L. Pasquali and M. H. Reno, 
  Phys.\ Rev.\  D {\bf 59}, 093003 (1999);
 A.~D.~Martin, M.~G.~Ryskin and A.~M.~Stasto,
  Acta Phys.\ Polon.\  B {\bf 34}, 3273 (2003).

\bibitem{leelin}
F.~F.~Lee and G.~L.~Lin,
  Astropart.\ Phys.\  {\bf 25}, 64 (2006).
\bibitem{gmm}
G.~Giordano, O.~Mena and I.~Mocioiu,
  arXiv:1004.3519 [hep-ph].
 
\bibitem{covi}
See, e.g.,
  L.~Covi, M.~Grefe, A.~Ibarra and D.~Tran,
  JCAP {\bf 0901}, 029 (2009).
  
\bibitem{albright}
 C.~H.~Albright and C.~Jarlskog,
  Nucl.\ Phys.\  B {\bf 84}, 467 (1975).
  
\bibitem{krtau}
 S.~Kretzer and M.~H.~Reno,
  Phys.\ Rev.\  D {\bf 66}, 113007 (2002).
\bibitem{proc}
See also,  S.~Kretzer and M.~H.~Reno,
  Nucl.\ Phys.\ Proc.\ Suppl.\  {\bf 139}, 134 (2005).

  
\bibitem{charm}
T.~Gottschalk,
  Phys.\ Rev.\  D {\bf 23}, 56 (1981);
       J.~J.~van der Bij and G.~J.~van Oldenborgh,
        Z.\ Phys.\  C {\bf 51}, 477 (1991);
 J.~Smith and W.~L.~van Neerven,
  Nucl.\ Phys.\  B {\bf 374}, 36 (1992);
 G.~Kramer and B.~Lampe,
  Z.\ Phys.\  C {\bf 54}, 139 (1992).

 
\bibitem{schienbein}
 I.~Schienbein {\it et al.},
  J.\ Phys.\ G {\bf 35}, 053101 (2008) and
  references therein.

\bibitem{kr}
 S.~Kretzer and M.~H.~Reno,
  Phys.\ Rev.\  D {\bf 69}, 034002 (2004).
  
\bibitem{georgi}
H. Georgi and H. D. Politzer, Phys. Rev. D {\bf 14},
1829 (1976); A. De Rujula, H. Georgi and H. D. Politzer,
Annals Phys. {\bf 103}, 315 (1977); R. Barbieri, J. R. Ellis,
M. K. Gaillard and G. G. Ross, Nucl. Phys. B {\bf 117}, 50 (1976). 




\bibitem{aot}
M.~A.~G.~Aivazis, F.~I.~Olness and W.~K.~Tung,
  Phys.\ Rev.\  D {\bf 50}, 3085 (1994).
   
\bibitem{acot}  
M.~A.~G.~Aivazis, F.~I.~Olness and W.~K.~Tung,
  Phys.\ Rev.\  D {\bf 50}, 3085 (1994);
   W.~K.~Tung, S.~Kretzer and C.~Schmidt,
  J.\ Phys.\ G {\bf 28}, 983 (2002).
   
\bibitem{leshouches}
 J.~R.~Andersen {\it et al.}  [SM and NLO Multileg Working Group],
  arXiv:1003.1241 [hep-ph].

  




\bibitem{llsmith}
C.~H.~Llewellyn Smith,
  Phys.\ Rept.\  {\bf 3}, 261 (1972).

\bibitem{sv}
 A.~Strumia and F.~Vissani,
  Phys.\ Lett.\  B {\bf 564}, 42 (2003).

\bibitem{fewpion}
D.~Rein and L.~M.~Sehgal,
  Annals Phys.\  {\bf 133}, 79 (1981).
G.~L.~Fogli and G.~Nardulli,
  Nucl.\ Phys.\  B {\bf 160}, 116 (1979).
  
\bibitem{fewpionnew}
See, e.g.,
K.~Hagiwara, K.~Mawatari and H.~Yokoya,
  Nucl.\ Phys.\  B {\bf 668}, 364 (2003)
  [Erratum-ibid.\  B {\bf 701}, 405 (2004)];
E.~A.~Paschos and J.~Y.~Yu,
  Phys.\ Rev.\  D {\bf 65}, 033002 (2002);
S.~Ahmad, M.~Sajjad Athar and S.~K.~Singh,
  Phys.\ Rev.\  D {\bf 74}, 073008 (2006).


\bibitem{ckmt}
A. Capella, A. Kaidalov, C. Merino and J. Tran Thanh Van, Phys. Lett. B {\bf 337},
358 (1994) and {\it Proceedings of 29th Rencontres de Moriond: QCD and high energy hadronic
interactions, Meribel les Allues, France, 19-26 March 1994}, pp. 271-282.

\bibitem{reno}
M.~H.~Reno,
  Phys.\ Rev.\  D {\bf 74}, 033001 (2006).
  
\bibitem{bodek} 
U. K. Yang and A. Bodek, Phys. Rev. 
Lett. {\bf 82}, 2467 (1999);
A.~Bodek and U.~K.~Yang,
  AIP Conf.\ Proc.\  {\bf 670}, 110 (2003);
  AIP Conf.\ Proc.\  {\bf 792}, 257 (2005).


\bibitem{gjr}
 M.~Gluck, P.~Jimenez-Delgado and E.~Reya,
  Eur.\ Phys.\ J.\  C {\bf 53}, 355 (2008).

\bibitem{cteq6p6}
P.~M.~Nadolsky {\it et al.},
  Phys.\ Rev.\  D {\bf 78}, 013004 (2008).

\bibitem{jr1}
  Y.~S.~Jeong and M.~H.~Reno,
  Phys. Rev. D {\bf 81}, 114012 (2010);
  M.~Gluck, P.~Jimenez-Delgado and E.~Reya,
  Phys. Rev. D {\bf 81}, 097501 (2010).
 


\bibitem{naumov}
 K.~S.~Kuzmin, V.~V.~Lyubushkin and V.~A.~Naumov,
  Phys.\ Atom.\ Nucl.\  {\bf 69}, 1857 (2006).

\bibitem{donut}
 K.~Kodama {\it et al.}  [DONuT Collaboration],
  Phys.\ Rev.\  D {\bf 78}, 052002 (2008).
  
\bibitem{opera}
N. Agafonova et al., OPERA collaboration, arXiv:1006.1623 [hep-ex].

\end{thebibliography}
\end{document}